\documentclass[12pt,a4paper]{amsart}
\usepackage{amsfonts}

\usepackage[latin1]{inputenc}
\usepackage{amsmath,amssymb,amsthm,bm}
\usepackage{graphics}

\theoremstyle{definition}

\def\<{\langle}
\def\>{\rangle}

\input{tcilatex}

\begin{document}
\title{Is the decoherence of a system the result of its interaction with the
environment?}
\author{Mario Castagnino}
\address{CONICET-IAFE-IFIR-Universidad de Buenos Aires.}
\author{Sebastian Fortin}
\address{CONICET-IAFE-Universidad de Buenos Aires. Email: sfortin@gmx.net}
\author{Olimpia Lombardi}
\address{CONICET-Universidad de Buenos Aires.}

\begin{abstract}
According to a usual reading, decoherence is a process resulting from the
interaction between a small system and its large environment where
information and energy are dissipated. The particular models treated in the
literature on the subject reinforce this idea since, in general, the
behavior of a particle immersed in a large ``bath'' composed by many
particles is studied. The aim of this letter is to warn against this usual
simplified reading. By means of the analysis of a well-known model, we will
show that decoherence may occur in a system interacting with an environment
consisting of \textit{only one particle}.
\end{abstract}

\maketitle

\paragraph{\textbf{Introduction.}}

The word \textquotedblleft decoherence\textquotedblright\ refers to the
quantum process that turns a coherent pure state into a decohered mixed
state, which is diagonal in a well defined basis. The phenomenon of
decoherence is essential in the account of the emergence of classicality
from quantum behavior, since it explains how interference vanishes in an
extremely short decoherence time.

The orthodox explanation of the phenomenon is given by the so-called
\textquotedblleft \textit{environment-induced decoherence\textquotedblright }
(EID) approach (\cite{Zurek-1982}, \cite{Zurek-1993}, \cite{Paz-Zurek}, \cite%
{Zurek-2003}), according to which decoherence is a process resulting from
the interaction of a quantum system and its environment. As Zurek states,
the environment destroys the coherence between the states of a quantum
system by its incessant \textquotedblleft monitoring\textquotedblright\ of
the observables associated with the preferred states: it is the environment
what \textquotedblleft distills\textquotedblright\ the classical essence
from quantum systems (see \cite{Paz-Zurek}, \cite{Zurek-2003}).\ In
addition, since decoherence only occurs in open quantum systems, it must
always be accompanied by other manifestations of openness, such as
dissipation of energy and information into the environment.

This way of presenting decoherence has led to a standard reading of the
physical meaning of the phenomenon. According to this reading, decoherence
is a process resulting from the interaction between a small system and its
large environment, where information and energy are dissipated. The
particular models treated in the literature on the subject have reinforced
this idea since, in general, the behavior of a particle immersed in a large
\textquotedblleft bath\textquotedblright\ composed by many particles is
studied. The aim of this letter is to warn against this usual simplified
reading. \ By means of the analysis of a well-known model, we will show that
decoherence occurs in cases that cannot be described as a small system
interacting with a large environment.

\paragraph{\textbf{The spin-bath model.}}

The spin-bath model is a very simple model that has been exactly solved in
previous papers (see \cite{Zurek-1982}). We will study it from the general
theoretical framework for decoherence presented in a previous work \cite%
{CQG-CFLL-08}. Let us consider a closed system $U=S+E$ where (i) the system $%
S$ is a spin-1/2 particle $P$ represented in the Hilbert space $\mathcal{H}%
_{S}$, and (ii) the environment $E$ is composed of $N$ spin-1/2 particles $%
P_{i}$, each one of which is represented in its own Hilbert space $\mathcal{H%
}_{i}$. The complete Hilbert space of the composite system $U$\ is, $%
\mathcal{H}=\mathcal{H}_{S}\bigotimes\limits_{i=1}^{N}\mathcal{H}_{i}$. In
the particle $P$, the two eigenstates of the spin operator $S_{S,%
\overrightarrow{v}}$\ in direction $\overrightarrow{v}$ are $\left\vert
\Uparrow \right\rangle $ and $\left\vert \Downarrow \right\rangle $, such
that $S_{S,\overrightarrow{v}}\left\vert \Uparrow \right\rangle =\frac{1}{2}%
\left\vert \Uparrow \right\rangle $\ and\ $S_{S,\overrightarrow{v}%
}\left\vert \Downarrow \right\rangle =-\frac{1}{2}\left\vert \Downarrow
\right\rangle $. In each particle $P_{i}$, the two eigenstates of the
corresponding spin operator $S_{i,\overrightarrow{v}}$\ in direction $%
\overrightarrow{v}$ are $\left\vert \uparrow _{i}\right\rangle $ and $%
\left\vert \downarrow _{i}\right\rangle $, such that $S_{i,\overrightarrow{v}%
}\left\vert \uparrow _{i}\right\rangle =\frac{1}{2}\left\vert \uparrow
_{i}\right\rangle $ and\ $S_{i,\overrightarrow{v}}\left\vert \downarrow
_{i}\right\rangle =\frac{1}{2}\left\vert \downarrow _{i}\right\rangle $.
Therefore, a pure initial state of $U$ reads%
\begin{equation}
|\psi _{0}\rangle =(a\left\vert \Uparrow \right\rangle +b\left\vert
\Downarrow \right\rangle )\bigotimes_{i=1}^{N}(\alpha _{i}|\uparrow
_{i}\rangle +\beta _{i}|\downarrow _{i}\rangle )  \label{3.3}
\end{equation}%
where the coefficients $a$, $b$, $\alpha _{i}$, $\beta _{i}$ are such that
satisfy $\left\vert a\right\vert ^{2}+\left\vert b\right\vert ^{2}=1$ and $%
\left\vert \alpha _{i}\right\vert ^{2}+\left\vert \beta _{i}\right\vert
^{2}=1$. Usually these numbers (and also the $g_{i}$ below) are taken as
aleatory numbers. The self-Hamiltonians $H_{S}$ and $H_{E}$ of $S$ and $E$,
respectively, are taken to be zero, then the total Hamiltonian $%
H=H_{S}+H_{E}+H_{SE}$ of the composite system $U$ results (see \cite%
{Zurek-1982}, \cite{Max})%
\begin{equation}
H=H_{SE}=S_{S,\overrightarrow{v}}\otimes \sum_{i=1}^{N}2g_{i}S_{i,%
\overrightarrow{v}}\bigotimes_{j\neq i}^{N}I_{j}  \label{3.4}
\end{equation}%
where $I_{j}$ is the identity operator on the subspace $\mathcal{H}_{j}$, $%
S_{S,\overrightarrow{v}}=\frac{1}{2}\left( \left\vert \Uparrow \right\rangle
\left\langle \Uparrow \right\vert -\left\vert \Downarrow \right\rangle
\left\langle \Downarrow \right\vert \right) $ and $S_{i,\overrightarrow{v}}=%
\frac{1}{2}\left( \left\vert \uparrow _{i}\right\rangle \left\langle
\uparrow _{i}\right\vert -\left\vert \downarrow _{i}\right\rangle
\left\langle \downarrow _{i}\right\vert \right) $. Under the action of $%
H=H_{SE}$, the state $|\psi _{0}\rangle $ evolves as $\left\vert \psi
(t)\right\rangle =a\left\vert \Uparrow \right\rangle |\mathcal{E}_{\Uparrow
}(t)\rangle +b\left\vert \Downarrow \right\rangle |\mathcal{E}_{\Downarrow
}(t)\rangle $ where $\left\vert \mathcal{E}_{\Uparrow }(t)\right\rangle
=\left\vert \mathcal{E}_{\Downarrow }(-t)\right\rangle $ and%
\begin{equation}
\left\vert \mathcal{E}_{\Uparrow }(t)\right\rangle
=\bigotimes_{i=1}^{N}\left( \alpha _{i}\,e^{ig_{i}t/2}\,\left\vert \uparrow
_{i}\right\rangle +\beta _{i}\,e^{-ig_{i}t/2}\,\left\vert \downarrow
_{i}\right\rangle \right)  \label{3.6}
\end{equation}

If $\mathcal{O}$ is the space of observables of the whole system $U$, let us
consider a space of relevant observables $\mathcal{O}_{R}\subset \mathcal{O}$
such that $O_{R}\in \mathcal{O}_{R}$ reads%
\begin{equation}
O_{R}=\left( 
\begin{array}{c}
s_{\Uparrow \Uparrow }\left\vert \Uparrow \right\rangle \left\langle
\Uparrow \right\vert \\ 
+s_{\Uparrow \Downarrow }\left\vert \Uparrow \right\rangle \left\langle
\Downarrow \right\vert \\ 
+s_{\Downarrow \Uparrow }\left\vert \Downarrow \right\rangle \left\langle
\Uparrow \right\vert \\ 
+s_{\Downarrow \Downarrow }\left\vert \Downarrow \right\rangle \left\langle
\Downarrow \right\vert%
\end{array}%
\right) \bigotimes_{i=1}^{N}\left( 
\begin{array}{c}
\epsilon _{\uparrow \uparrow }^{(i)}\left\vert \uparrow _{i}\right\rangle
\left\langle \uparrow _{i}\right\vert \\ 
+\epsilon _{\downarrow \downarrow }^{(i)}\left\vert \downarrow
_{i}\right\rangle \left\langle \downarrow _{i}\right\vert \\ 
+\epsilon _{\downarrow \uparrow }^{(i)}\left\vert \downarrow
_{i}\right\rangle \left\langle \uparrow _{i}\right\vert \\ 
+\epsilon _{\uparrow \downarrow }^{(i)}\left\vert \uparrow _{i}\right\rangle
\left\langle \downarrow _{i}\right\vert%
\end{array}%
\right)  \label{3.7}
\end{equation}%
Since the operators $O_{R}$ are Hermitian, the diagonal components $%
s_{\Uparrow \Uparrow }$, $s_{\Downarrow \Downarrow }$, $\epsilon _{\uparrow
\uparrow }^{(i)}$,$\epsilon _{\downarrow \downarrow }^{(i)}$ are real
numbers and the off-diagonal components are complex numbers satisfying $%
s_{\Uparrow \Downarrow }=s_{\Downarrow \Uparrow }^{\ast }$, $\epsilon
_{\uparrow \downarrow }^{(i)}=\epsilon _{\downarrow \uparrow }^{(i)\ast }$.
Then, the expectation value of the observable $O$ in the state $\left\vert
\psi (t)\right\rangle $ can be computed as%
\begin{eqnarray}
\langle O_{R}\rangle _{\psi (t)} &=&(|a|^{2}s_{\Uparrow \Uparrow
}+|b|^{2}s_{\Downarrow \Downarrow })\,\Gamma _{0}(t)  \notag \\
&&+2\func{Re}\,[ab^{\ast }\,s_{\Downarrow \Uparrow }\,\Gamma _{1}(t)]
\label{3.8}
\end{eqnarray}%
where (see eqs. (23) and (24) in \cite{Max}) 
\begin{align}
\Gamma _{0}(t)& =\prod_{i=1}^{N}\left[ 
\begin{array}{c}
|\alpha _{i}|^{2}\epsilon _{\uparrow \uparrow }^{(i)}+\alpha _{i}{}^{\ast
}\beta _{i}\epsilon _{\uparrow \downarrow }^{(i)}e^{-ig_{i}t} \\ 
+|\beta _{i}|^{2}\epsilon _{\downarrow \downarrow }^{(i)}+(\alpha
_{i}{}^{\ast }\beta _{i}\epsilon _{\uparrow \downarrow }^{(i)})^{\ast
}e^{ig_{i}t}%
\end{array}%
\right]  \label{3.9} \\
\Gamma _{1}(t)& =\prod_{i=1}^{N}\left[ 
\begin{array}{c}
|\alpha _{i}|^{2}\epsilon _{\uparrow \uparrow }^{(i)}e^{ig_{i}t}+|\beta
_{i}|^{2}\epsilon _{\downarrow \downarrow }^{(i)}e^{-ig_{i}t} \\ 
+\alpha _{i}{}^{\ast }\beta _{i}\epsilon _{\uparrow \downarrow
}^{(i)}+(\alpha _{i}{}^{\ast }\beta _{i}\epsilon _{\uparrow \downarrow
}^{(i)})^{\ast }%
\end{array}%
\right]  \label{3.10}
\end{align}

As a generalization of the usual presentations, we will study different ways
of splitting the whole closed system $U$ into a relevant part and its
environment, by considering different choices for the space $\mathcal{O}_{R}$%
.

\paragraph{\textbf{Case 1: A large environment that produces decoherence.}}

In the typical situation studied by the EID approach, the system of interest 
$S$ is simply the particle $P$. Therefore, the relevant observables $%
O_{R}\in \mathcal{O}_{R}$ are those corresponding to $P$, and are obtained
from eq. (\ref{3.7}) by making $\epsilon _{\uparrow \uparrow
}^{(i)}=\epsilon _{\downarrow \downarrow }^{(i)}=1,$ $\epsilon _{\uparrow
\downarrow }^{(i)}=0$:%
\begin{equation}
O_{R}=\left( \sum_{s,s^{\prime }=\Uparrow ,\Downarrow }s_{ss^{\prime
}}|s\rangle \langle s^{\prime }|\right)
\bigotimes_{i=1}^{N}I_{i}=O_{S}\bigotimes_{i=1}^{N}I_{i}  \label{3.11}
\end{equation}%
The expectation value of these observables is given by%
\begin{equation}
\langle O_{R}\rangle _{\psi (t)}=|a|^{2}\,s_{\Uparrow \Uparrow
}+|b|^{2}\,s_{\Downarrow \Downarrow }+2\func{Re}[ab^{\ast }\,s_{\Downarrow
\Uparrow }\,r_{1}(t)]  \label{3.12}
\end{equation}%
where 
\begin{equation}
r_{1}(t)=\prod_{i=1}^{N}\left[ |\alpha _{i}|^{2}e^{ig_{i}t}+|\beta
_{i}|^{2}e^{-ig_{i}t}\right]  \label{3.13}
\end{equation}%
By comparing eq. (\ref{3.12}) with eq. (\ref{3.8}), we see that in this case 
$\Gamma _{0}(t)=1$ and $\Gamma _{1}(t)=r_{1}(t)$. Moreover,

\begin{equation}
|r_{1}(t)|^{2}=\prod_{i=1}^{N}(|\alpha _{i}|^{4}+|\beta _{i}|^{4}+2|\alpha
_{i}|^{2}|\beta _{i}|^{2}\cos 2g_{i}t)  \label{3.14}
\end{equation}%
Since $|\alpha _{i}|^{2}+|\beta _{i}|^{2}=1$, then 
\begin{align}
& \max_{t}(|\alpha _{i}|^{4}+|\beta _{i}|^{4}+2|\alpha _{i}|^{2}|\beta
_{i}|^{2}\cos 2g_{i}t)  \notag \\
& =\left( \left( |\alpha _{i}|^{2}+|\beta _{i}|^{2}\right) ^{2}\right) =1
\label{3.15}
\end{align}%
and%
\begin{eqnarray}
&&\min_{t}\left( \left\vert \alpha _{i}\right\vert ^{4}+\left\vert \beta
_{i}\right\vert ^{4}+2\left\vert \alpha _{i}\right\vert ^{2}\left\vert \beta
_{i}\right\vert ^{2}\cos \left( 2g_{i}t\right) \right)  \notag \\
&=&\left( \left( |\alpha _{i}|^{2}-|\beta _{i}|^{2}\right) ^{2}\right)
=\left( 2\left\vert \alpha _{i}\right\vert ^{2}-1\right) ^{2}  \label{3.15'}
\end{eqnarray}%
If the coefficients $g_{i}$, $\alpha _{i}$ and $\beta _{i}$ are aleatory
numbers, then $(|\alpha _{i}|^{4}+|\beta _{i}|^{4}+2|\alpha _{i}|^{2}|\beta
_{i}|^{2}\cos 2g_{i}t)$ is an aleatory number which, if $t\neq 0$,
fluctuates between $1$ and $\left( 2\left\vert \alpha _{i}\right\vert
^{2}-1\right) ^{2}$. Let us note that, since the $\left\vert \alpha
_{i}\right\vert ^{2}$ and the $\left\vert \beta _{i}\right\vert ^{2}$ are
aleatory numbers in the closed interval $\left[ 0,1\right] $, when the
environment has many particles (that is, when $N\rightarrow \infty $), the
statistical value of the cases $\left\vert \alpha _{i}\right\vert ^{2}=1$, $%
\left\vert \beta _{i}\right\vert ^{2}=1$, $\left\vert \alpha _{i}\right\vert
^{2}=0$ and $\left\vert \beta _{i}\right\vert ^{2}=0$ is zero. In this case,
eq. (\ref{3.14}) for $|r_{1}(t)|^{2}$ is an infinite product of numbers
belonging to the open interval $\left( 0,1\right) $. \ As a consequence (see 
\cite{Paz-Zurek}, \cite{Zurek-2003}), 
\begin{equation}
\lim_{N\rightarrow \infty }r_{1}(t)=0  \label{3.16}
\end{equation}

In order to know the time-behavior of the expectation value of eq. (\ref%
{3.12}), we have to compute the time-behavior of $r_{1}(t)$. If we know that 
$r_{1}(0)=1$ for $N\rightarrow \infty $, and that $\lim_{N\rightarrow \infty
}r_{1}(t)=0$ for any $t\neq 0$, it can be expected that, for $N$ finite, $%
r_{1}(t)$ will evolve in time from $r_{1}(0)=1$ to a very small value. \
Moreover, $r_{1}(t)$ is a periodic function because it is a product of
periodic functions with periods depending on the coefficients $g_{i}$.
Nevertheless, since the $g_{i}$ are aleatory, the periods of the individual
functions are different and, as a consequence, the recurrence time of $%
r_{1}(t)$ will be very large, and strongly increasing with the number $N$ of
particles.

The time-behavior of $r_{1}(t)$ was computed by means of a numerical
simulation, where the aleatory numbers $\left\vert \alpha _{i}\right\vert
^{2}$, $\left\vert \beta _{i}\right\vert ^{2}$\ and $g_{i}$\ were obtained
from a generator of aleatory numbers: these generator fixed the value of $%
\left\vert \alpha _{i}\right\vert ^{2}$, and the $\left\vert \beta
_{i}\right\vert ^{2}$ were computed as $\left\vert \beta _{i}\right\vert
^{2}=1-\left\vert \alpha _{i}\right\vert ^{2}$. The function $r_{1}(t)$ for $%
N=200$ is plotted in Figure \ref{fig 1}, which shows that the particle $P$
decoheres in interaction with an environment of $N$ particles $P_{i}$. This
result (see also numerical simulations in \cite{Max}) agrees with the
standard reading of the phenomenon of decoherence: a single particle in
interaction with a large environment of many particles decoheres due
precisely to that interaction.

\begin{figure}[t]
\par
\centerline{\scalebox{0.7}{\includegraphics{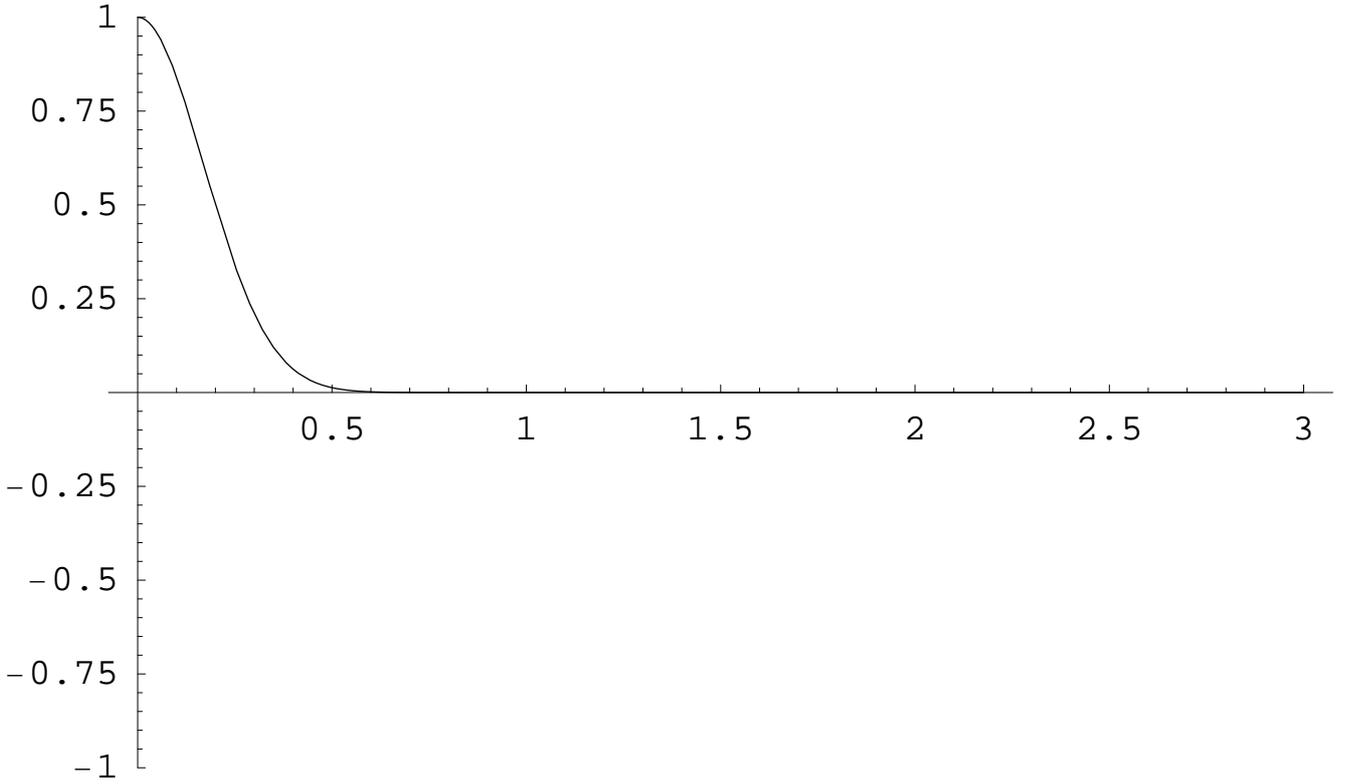}}} 
\vspace*{0.cm}
\caption{Plot of $|r_{1}(t)|^{2}$ given by eq. (\protect\ref{3.14}),
for $N=200$.}
\label{f1}
\end{figure}

\paragraph{\textbf{Case 2: A large environment with no decoherence.}}

Although in the usual presentations of the model the system of interest is $%
P $, as in the previous section, we can conceive different ways of splitting
the whole system $U$ into a system of interest and an environment. For
instance, it may be the case that the measuring arrangement
\textquotedblleft observes\textquotedblright\ a particular particle $P_{j}$
of what was previously considered the environment. In this case, the system
of interest $S$ is the particle $P_{j}$, and the environment is composed by
all the remaining particles, $E=P+\sum_{i\neq j}P_{i}$. Then, the relevant
observables $O_{R_{j}}\in \mathcal{O}_{R_{j}}\subset \mathcal{O}$ are only
those corresponding to $P_{j}$: $O_{R_{j}}=I_{S}\otimes
O_{S_{j}}\bigotimes_{i\neq j}I_{i}$ where%
\begin{eqnarray}
O_{S_{j}} &=&\epsilon _{\uparrow \uparrow }^{(j)}\,|\uparrow _{j}\rangle
\langle \uparrow _{j}|+\epsilon _{\downarrow \downarrow }^{(j)}\,|\downarrow
_{j}\rangle \langle \downarrow _{j}|  \notag \\
&&+\epsilon _{\downarrow \uparrow }^{(j)}\,|\downarrow _{j}\rangle \langle
\uparrow _{j}|+\epsilon _{\uparrow \downarrow }^{(j)}\,|\uparrow _{j}\rangle
\langle \downarrow _{j}|  \label{4-1.2}
\end{eqnarray}
where the coefficients $\epsilon _{\uparrow \uparrow }^{(j)}$, $\epsilon
_{\downarrow \downarrow }^{(j)}$, $\epsilon _{\downarrow \uparrow }^{(j)}$
are now generic. The expectation value of the observables $O_{R_{j}}$ is
given by%
\begin{eqnarray}
\langle O_{R_{j}}\rangle _{\psi (t)} &=&\left\vert \alpha _{j}\right\vert
^{2}\epsilon _{\uparrow \uparrow }^{(j)}+\left\vert \beta _{j}\right\vert
^{2}\epsilon _{\downarrow \downarrow }^{(j)}  \notag \\
&&+\func{Re}\left( \alpha _{j}\beta _{j}^{\ast }\epsilon _{\uparrow
\downarrow }^{(j)}e^{ig_{j}t}\right)  \label{4-1.3}
\end{eqnarray}

In order to know the time-evolution of the expectation value of the $%
O_{R_{j}}$, we have to compute the time-behavior of the third term of eq. (%
\ref{4-1.3}):%
\begin{equation}
r_{2}(t)=\func{Re}\left( \alpha _{j}\beta _{j}^{\ast }\epsilon _{\uparrow
\downarrow }^{(j)}e^{ig_{j}t}\right)  \label{4-1.4}
\end{equation}%
Let us note that this equation is independent of $N\geq 1$. In this case,
numerical simulations are not required to see that $r_{2}(t)$ is an
oscillating function which, as a consequence, has no limit for $t\rightarrow
\infty $. Nevertheless, in order to illustrate the non decoherence of the
system $S$ we show the time-evolution of $r_{2}(t)$ with $N\geq 1$ in Figure %
\ref{fig 2}. In this case, a single particle $S=P_{j}$ with a large
environment $E=P+\sum_{i\neq j}P_{i}$ of $N$ particles does not decohere.
Nevertheless, this result can be accommodated under the standard reading of
the phenomenon of decoherence by saying that $P_{j}$ strongly interacts only
with particle $P$, but does not interact with the rest of the particles $%
P_{i\neq j}$; therefore, the interaction of $S=P_{j}$ with its environment $%
E=P+\sum_{i\neq j}P_{i}$ is not strong enough to produce decoherence.

\begin{figure}[t]
\par
\centerline{\scalebox{0.7}{\includegraphics{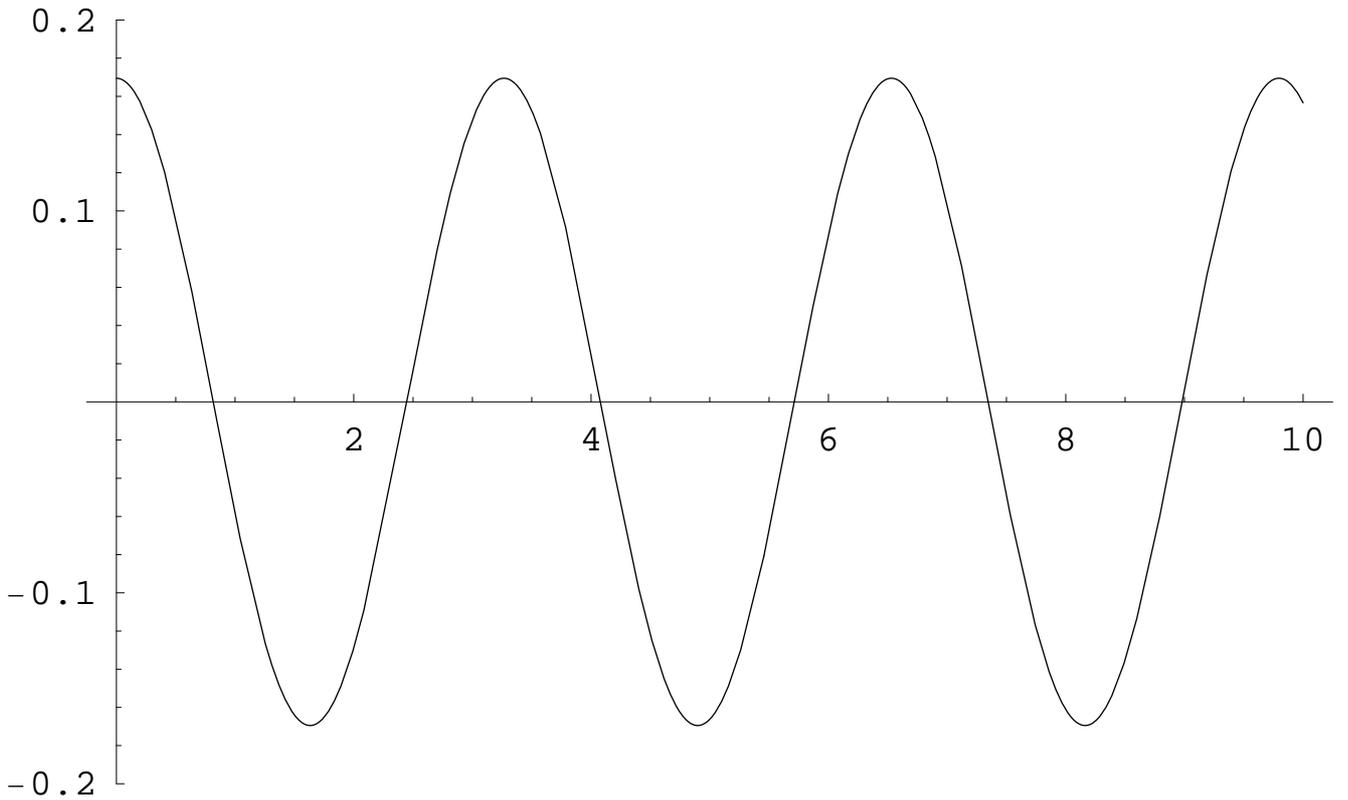}}} 
\vspace*{0.cm}
\caption{Plot of $r_{2}(t)$ given by eq. (\protect\ref{4-1.4}), for $N\geq 1$.}
\label{f2}
\end{figure}

\paragraph{\textbf{Case 3: A small environment that produces decoherence.}}

In this section we consider a measuring arrangement that \textquotedblleft
observes\textquotedblright\ a set of particles of the environment, e.g., the 
$p$ first particles $P_{j}$. In this case, the system of interest is
composed by $p$ particles, $S=$ $\sum\limits_{i=1}^{p}P_{i}$, and the
environment is composed by all the remaining particles, $E=P+\sum%
\limits_{i=p+1}^{N}P_{i}$. \ So, in eq. (\ref{3.7}), $s_{\Uparrow \Uparrow
}=s_{\Downarrow \Downarrow }=1$, $s_{\Uparrow \Downarrow }=s_{\Downarrow
\Uparrow }=0$, the coefficients $\epsilon _{\uparrow \uparrow }^{(j)}$, $%
\epsilon _{\downarrow \downarrow }^{(j)}$, $\epsilon _{\downarrow \uparrow
}^{(j)}$ are generic for $j\in \left\{ 1...p\right\} $, and $\epsilon
_{\uparrow \uparrow }^{(i)}=\epsilon _{\downarrow \downarrow }^{(i)}=1$, $%
\epsilon _{\downarrow \uparrow }^{(i)}=\epsilon _{\uparrow \downarrow
}^{(i)}=0$ for $i\in \left\{ p+1...N\right\} $. Then, the relevant
observables $O_{R}\in \mathcal{O}_{R}\subset \mathcal{O}$\ read%
\begin{equation}
O_{R}=I_{S}\otimes \left( \bigotimes_{j=1}^{p}O_{S_{j}}\right) \otimes
\left( \bigotimes_{i=p+1}^{N}I_{i}\right)  \label{5-1.1}
\end{equation}%
where $O_{S_{j}}$ is given by eq. (\ref{4-1.2}). Therefore, the expectation
value of the relevant observables $O_{R}$ is 
\begin{equation}
\langle O_{R}\rangle _{\psi (t)}=\prod_{i=1}^{p}\left[ 
\begin{array}{c}
|\alpha _{i}|^{2}\epsilon _{\uparrow \uparrow }^{(i)}+\alpha _{i}{}^{\ast
}\beta _{i}\epsilon _{\uparrow \downarrow }^{(i)}e^{-ig_{i}t} \\ 
+|\beta _{i}|^{2}\epsilon _{\downarrow \downarrow }^{(i)}+(\alpha
_{i}{}^{\ast }\beta _{i}\epsilon _{\uparrow \downarrow }^{(i)})^{\ast
}e^{ig_{i}t}%
\end{array}%
\right]  \label{5-1.4}
\end{equation}

Although eq. (\ref{5-1.4}) is very similar to eq. (\ref{3.10}), we will
compute the time-behavior of that expectation value by means of numerical
simulations.\ In order to simplify the computation, we will consider the
particular case where the relevant observables are

\begin{equation}
O_{R}=I_{S}\otimes \left( \bigotimes_{j=1}^{p}S_{x}^{(j)}\right) \otimes
\left( \bigotimes_{i=p+1}^{N}I_{i}\right)  \label{6-1.1}
\end{equation}%
where $S_{x}^{(j)}$ is the projection of the spin onto the $x$-axis of the
particle $P_{j}$. Then, $\epsilon _{\uparrow \uparrow }^{(j)}=\epsilon
_{\downarrow \downarrow }^{(j)}=0$, and the expectation value reads 
\begin{equation}
\langle O_{R}\rangle _{\psi (t)}=r_{3}(t)=\prod_{i=1}^{p}\left[ 2\ast \func{%
Re}\left( \alpha _{i}{}^{\ast }\beta _{i}\epsilon _{\uparrow \downarrow
}^{(i)}e^{-ig_{i}t}\right) \right]  \label{6-1.2}
\end{equation}%
As in eq. (\ref{4-1.4}), in this equation we can select any $N\geq P$. As in
Case 1 (see eq. (\ref{3.13})), in this case the time-dependence of $r_{3}(t)$
is given by a periodic function, whose recurrence time strongly increases
with the number of the involved particles.

\begin{figure}[t]
\par
\centerline{\scalebox{0.7}{\includegraphics{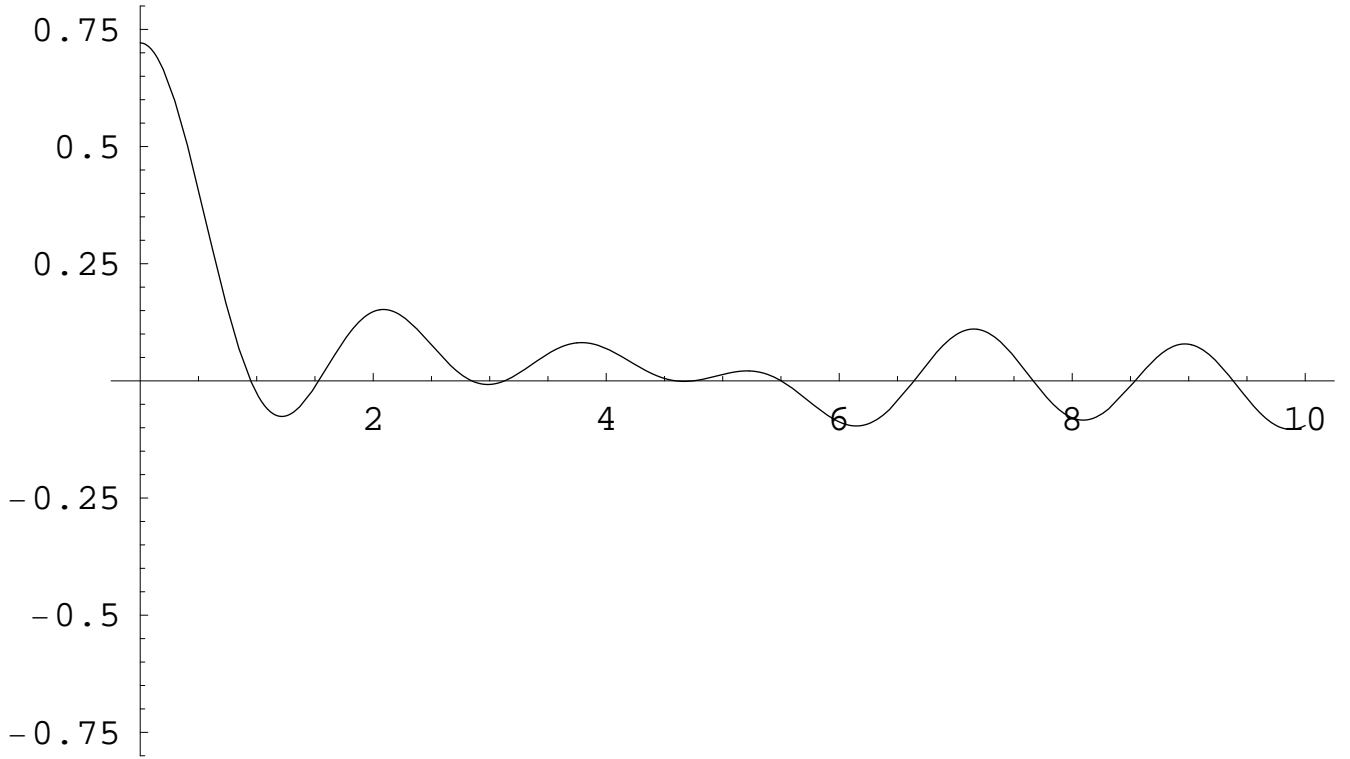}}} 
\vspace*{0.cm}
\caption{Plot of $r_{3}(t)$ given by eq. (\protect\ref{6-1.2}), for $%
p=4 $.}
\label{f3}
\end{figure}

The time-behavior of $r_{3}(t)$, with $p=4$, is plotted in Figure \ref{fig 3}%
, where we can see a fast decaying followed by fluctuations around zero. As
expected, such fluctuations strongly damp off with the increase of the
number $p$ of particles, as shown in Figure \ref{fig 4} ($p=8$) and Figure 5
($p=10$); with $p=200$ the plot turns out to be indistinguishable of that
obtained for the decoherence of Case 1 with $N=200$.

\begin{figure}[t]
\par
\centerline{\scalebox{0.7}{\includegraphics{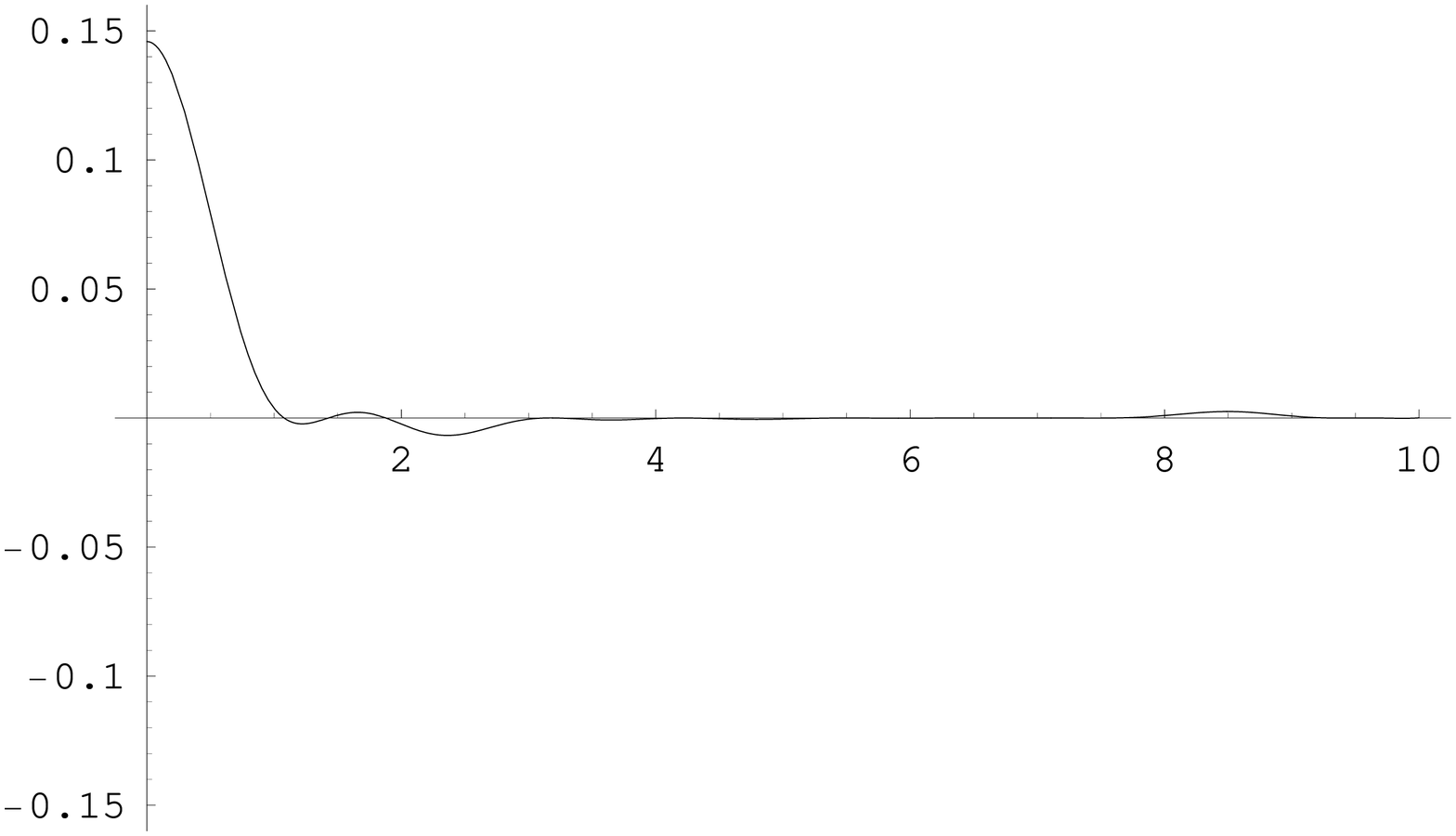}}} 
\vspace*{0.cm}
\caption{Plot of $r_{3}(t)$ given by eq. (\protect\ref{6-1.2}), for $%
p=8 $.}
\label{f4}
\end{figure}

The surprising consequence of these results is that the time-behavior is
independent of the number $N$ of the particles $P_{i}$, but only depends on
the number $p$ of the particles that constitute the system of interest (see
eq. (\ref{5-1.4})). Therefore, we can consider a limit case of $N=p=10$,
where the system $S$ is composed by the $p=N=10$ particles and the
environment $E$ is a single particle, $E=P$: in this case, as shown in
Figure \ref{fig 5}, we have to say that a system of $10$ particles decoheres
as the result of its interaction with a \textit{single-particle environment}%
.\ The situation becomes even more striking as the number $p$ increases:
with $N=p=200$, the system of $200$ particles strongly decoheres in
interaction with a single-particle environment. These results can hardly be
accommodated under the standard reading of the phenomenon of decoherence,
according to which decoherence is produced by the interaction between a
small system and a \textit{large environment}. In other words, this result
is in complete contradiction with the usual intuition behind EID.

\begin{figure}[t]
\par
\centerline{\scalebox{0.7}{\includegraphics{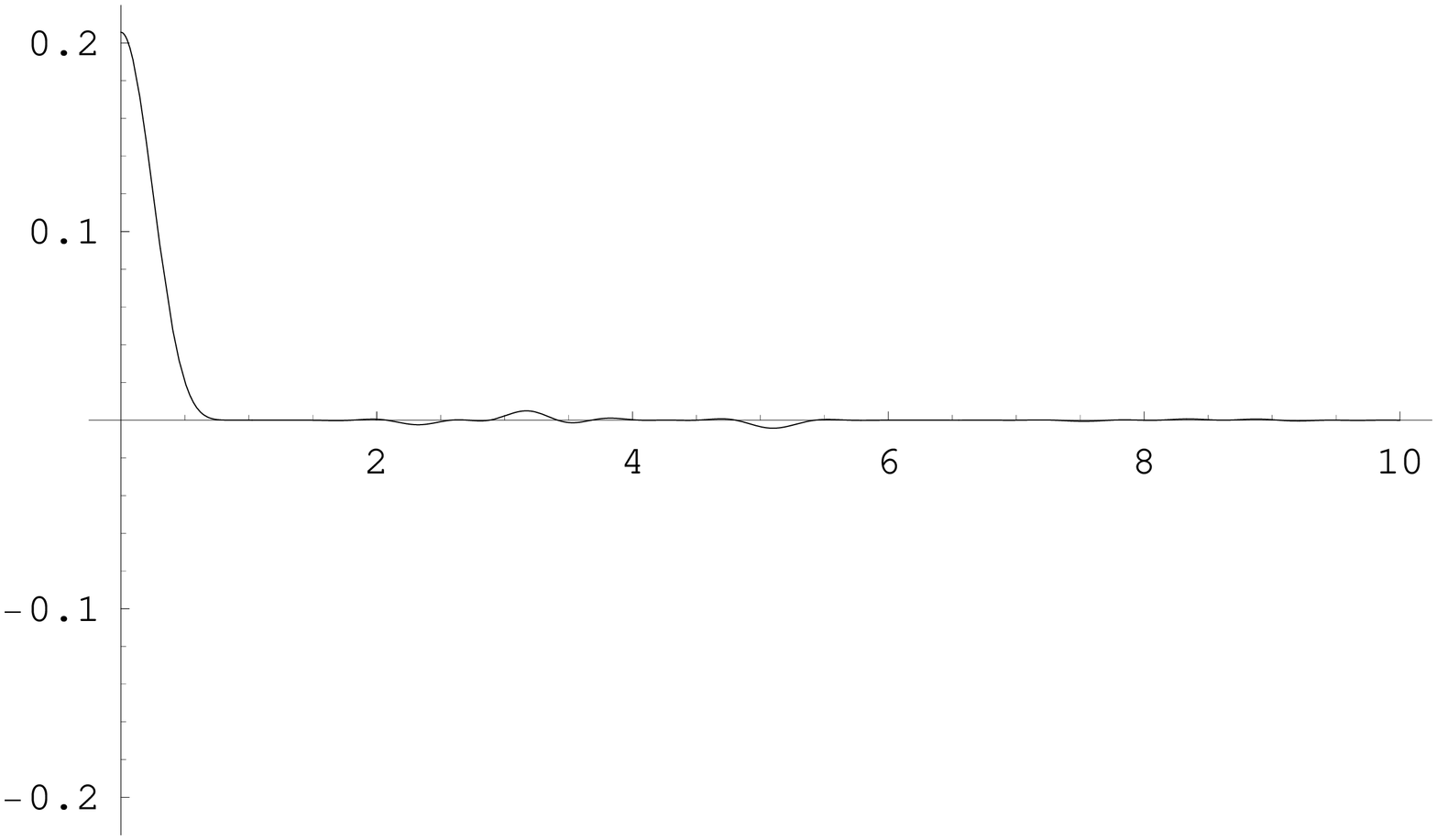}}} 
\vspace*{0.cm}
\caption{Plot of $r_{3}(t)$ given by eq. (\protect\ref{6-1.2}), for $%
p=10$.}
\label{f5}
\end{figure}

\paragraph{\textbf{Conclusions.}}

As some authors point out, the theory of decoherence has became the
\textquotedblleft new orthodoxy\textquotedblright\ in the quantum physicists
community (see \cite{Bub}). At present, decoherence is studied and tested in
many areas such as atomic physics, quantum optics and condensed matter, and
it has acquired a great relevance in quantum computation. This impressive
success has led to forget the questions about the physical meaning of
decoherence. In general, decoherence is expected to occur only when a small
system interacts with a large environment: the dissipation of information
and energy from the system to the large environment is what should cause the
destruction of the coherence between the states of the system.

By studying a well-known model from different perspectives, in this letter
we have shown that the usual way of understanding the physical meaning of
decoherence is, at least, misguided: a large system in interaction with a
small environment may decohere under particular conditions. The general
moral of this work is that our understanding of the conceptual foundations
of the phenomenon of decoherence is still far from being satisfactory, and
the matter deserves to be considered in detail by the physical community.

\paragraph{\textbf{Acknowledgments.}}

We are very grateful to Roland Omnès and Maximilian Schlösshauer for many
comments and criticisms. This research was partially supported by grants of
the University of Buenos Aires, CONICET and FONCYT of Argentina.

\end{document}